\documentclass[letterpaper, 10 pt, conference]{ieeeconf}  

\IEEEoverridecommandlockouts                              
\usepackage{amsmath}
\usepackage{multirow}
\usepackage[dvipsnames]{xcolor}
\usepackage{graphicx}
\usepackage{subcaption}
\usepackage{tabularx}
\usepackage{amsfonts}
\usepackage{url}

\title{\LARGE \bf
Learning Eco-Driving Strategies at Signalized Intersections
}

\author{Vindula Jayawardana$^{1}$ and Cathy Wu$^{2}$
\thanks{*This work was supported by the MIT-IBM Watson AI Lab.}
\thanks{$^{1}$Vindula Jayawardana is with MIT Laboratory for Information \& Decision Systems and Department of Electrical Engineering and Computer Science, Cambridge, MA, 02139, USA
        {\tt\small vindula@mit.edu}}%
\thanks{$^{2}$Cathy Wu is with Laboratory for Information \& Decision Systems, Department of Civil and Environmental Engineering and Institute of Data Systems and Society, Cambridge, MA, 02139, USA
        {\tt\small cathywu@mit.edu}}%
}

\begin{document}

\maketitle
\thispagestyle{empty}
\pagestyle{empty}

\begin{abstract}
Signalized intersections in arterial roads result in persistent vehicle idling and excess accelerations, contributing to fuel consumption and CO$_2$ emissions. There has thus been a line of work studying eco-driving control strategies to reduce fuel consumption and emission levels at intersections. However, methods to devise effective control strategies across a variety of traffic settings remain elusive. In this paper, we propose a reinforcement learning (RL) approach to learn effective eco-driving control strategies. We analyze the potential impact of a learned strategy on fuel consumption, CO$_2$ emission, and travel time and compare with naturalistic driving and model-based baselines. We further demonstrate the generalizability of the learned policies under mixed traffic scenarios. Simulation results indicate that scenarios with 100\% penetration of connected autonomous vehicles (CAV) may yield as high as 18\% reduction in fuel consumption and 25\% reduction in CO$_2$ emission levels while even improving travel speed by 20\%. Furthermore, results indicate that even 25\% CAV penetration can bring at least 50\% of the total fuel and emission reduction benefits. 


\end{abstract}

\section{Introduction}

Global greenhouse gas emission and fuel consumption rates have significantly increased over the last decade, indicating early signs of a major climate crisis in the near future. While many sectors actively contribute to greenhouse gas emission (GHG), transportation sector in the US contributes 29\% of which 77\% is due to land transportation~\cite{energy-literacy}. With clear indications that frequent vehicle stops created by stop-and-go waves~\cite{emission-waves, real-world-emission}, slow speeds on congested roads, over speeds~\cite{real-world-emission, article}, and idling can significantly increase fuel consumption and emission levels, carefully designed driving strategies are becoming imperative. The urge to reduce fuel consumption and related emission levels in driving has thus created a line of work on studying eco-driving strategies. 

In particular, eco-driving at signalized intersections has received significant attention due to the possible energy savings and reduction of emission levels. In arterial roads, traffic signals result in stop-and-go traffic waves producing acceleration, and idling events, increasing fuel consumption and emission levels. Recent studies have thus utilized connected automated vehicles (CAVs) as a means of control to achieve low fuel consumption and emissions when approaching and leaving an intersection~\cite{Yang2021EcoDrivingAS}. Such CAV control techniques fall under the broad umbrella of \textit{Lagrangian control}, which describes traffic control techniques based on mobile actuators (e.g. vehicles), rather than fixed-location actuators (e.g. traffic signals). In addition, the capability of CAVs to sense or communicate with nearby vehicles using V2V (Vehicle to Vehicle) and nearby intersections to exchange Signal and Phase Timing (SPaT) messages through V2I (Vehicle to Infrastructure) makes them well-suited control units for the Lagrangian control.


\begin{figure}
  \includegraphics[width=0.5\textwidth]{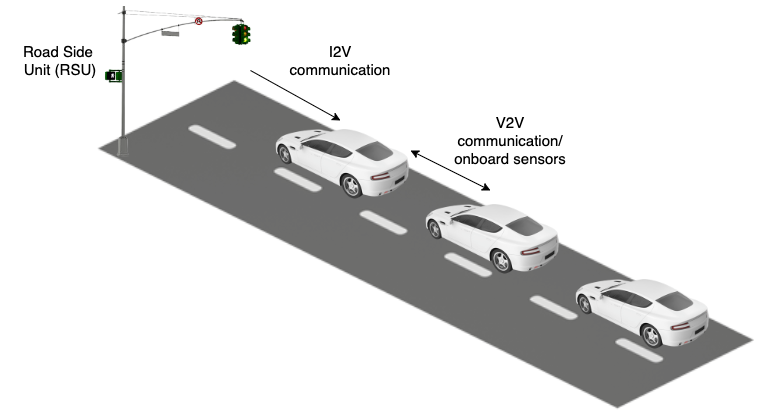} 
  \caption{Schematic overview of the approaching fleet of CAVs in a signalized intersection and the communication channels available for eco-driving}
  \label{fig:schematics}
\end{figure}

Most of the existing work on energy efficient driving at signalized intersections make assumptions on the vehicle dynamics and inter-vehicle dynamics, simplify the objective only to reduce fuel/emission without considering the impact on travel time, or usually involve solving a non-linear optimization problem in real time~\cite{Yang2021EcoDrivingAS, Cui2018ImpactOA,Ozkan2020APC}. Such methods fall under the broader class of model-based control methods.

Recently, learning-based model-free control methods have been proposed to solve challenging control problems. In particular, deep reinforcement learning has been used in obtaining desired control policies in many control tasks ranging from transportation~\cite{wu2017emergent, flow}, and healthcare~\cite{Yu2019ReinforcementLI} to economics and finance~\cite{RL-finance}. 
In this work, we leverage deep reinforcement learning (DRL) to obtain Lagrangian control policies for CAVs when approaching and leaving a signalized intersection. In particular, we take a multi-agent DRL approach.

Unlike most existing works, which focus on reducing fuel consumption or emission reduction without explicitly factoring in the impact on travel time caused by the new strategy, we formulate a more realistic objective by reducing fuel consumption while minimizing the impact on travel time. Since there is a general consensus that the fuel consumption is proportional to CO$_2$ emission levels~\cite{FONTARAS201797}, we also measure CO$_2$ emission levels under the resultant policy.

We show that the learned control policies perform on par with model-based baselines and significantly outperform naturalistic driving baselines, and the transfer capacity to out-of-distribution settings is successful.   

Our main contributions in this study are:

\begin{itemize}
    \item 
    We formulate Lagrangian control at intersections as a Partially Observable Markov Decision Process (POMDP) and solve it using reinforcement learning to reduce fuel consumption (and thereby emission levels) caused by accelerations, and idling while minimizing the impact on travel time.
    \item We assess the benefit of learned eco-driving control policies under naturalistic driving scenarios and model-based baselines and show significant savings in fuel and reductions in emission levels even while improving the average travel time. 
    \item We assess the generalizability of the learned policies by zero-shot transfer of the learned control policies to mixed traffic scenarios (varying CAV penetration rates) that were unseen at the training time.
\end{itemize}

\section{Related Work}

\subsection{Eco-driving}

Studies on eco-driving can be categorized into freeway-based and arterial-based control strategies. Freeway-based strategies are mainly concerned with maintaining a desirable speed without fluctuations, as traffic flow is rarely affected by traffic signals. This is usually achieved using advisory speed limits~\cite{eco-platoon} or platoon-based operations~\cite{BARTH2009400}. Arterial-based eco-driving is more involved due to the shock waves created by consecutive traffic signals. Yang et al. and Yang et al. introduce eco-driving control strategies based on queue length estimates and queue discharging estimates for single and multiple intersections, respectively~\cite{baseline-one,Yang2021EcoDrivingAS}. Ozkan et al. propose a receding-horizon optimization-based non-linear model predictive control algorithm for CAVs to reduce fuel consumption~\cite{Ozkan2020APC}. Zhao et al. propose a platoon-based eco-driving strategy for mixed traffic~\cite{ZHAO2018802}. Similarly, model predictive control (MPC) is employed in HomChaudhuri et al.~\cite{mpc-1} where as dynamic programming is used in Sun et al.~\cite{dp-1}. 

Additionally, there have been experimental field studies that focus on eco-driving. Many of them study the benefit of eco-driving that arises from smart driving after a fuel-efficient driving course was provided to a selected set of human drivers~\cite{BEUSEN2009514, ZARKADOULA2007449}. Further studies have also been carried out to field test and quantify the fuel saving benefits of eco-driving using CAVs~\cite{gov-eco-test, 8917077}.

However, all these work share a standard set of limitations. They usually assume a model of the dynamics, require solving a non-linear optimization problem in real-time, are not amenable to changes in the environments or make assumptions on queue lengths around the signalized intersection and the respective discharging rates. In contrast, our proposed DRL method does not assume any model of the dynamics, is faster to execute, and makes no assumptions of the queue length or discharging rates.   

\subsection{Reinforcement learning for autonomous traffic control}

Reinforcement learning has been gaining notable attention recently due to its ability to learn near-optimal control in complex environments. Many recent studies have attempted to use RL in devising control strategies in various traffic tasks. Wu et al. demonstrated the use of RL in mixed traffic scenarios and showed even a fraction of autonomous vehicles could stabilize the traffic from stop-and-go waves~\cite{flow}. Vinitsky et al. use RL to derive Lagrangian control policies for autonomous vehicles to improve the throughput of a bottleneck~\cite{bottleneck}. Yan et al.~\cite{Yan_2021} demonstrated that even without reward shaping, reinforcement learning based CAVs learn to coordinate to exhibit traffic signal-like behaviors~\cite{Yan_2021}. Many studies have utilized RL in traffic signal control and have shown significant benefits that are otherwise hard to obtain~\cite{intellight, DTlight}. 

Recently a few works have utilized RL to derive eco-driving control strategies. Zhu et al. use RL to seek optimal speed and power profiles to minimize fuel consumption over a given trip itinerary~\cite{Zhu2021ADR} for a single CAV. Similar work has been done by Guo et al. with an auxiliary model to factor in lane changing~\cite{GUO2021102980}. Wegner et al. propose a twin-delayed deep deterministic policy gradient method for eco-driving that takes only the information of the traffic light timing and minimal sensor data as inputs~\cite{WEGENER2021102967}. Although these works model multiple signalized intersections, they focus on reducing the fuel consumption for an ego agent; in contrast, our work seeks to optimize the full system of vehicles. Naturally, the prior work is then modeled as a single-agent problem; our work pursues a multi-agent formulation. 
 
\section{Preliminaries}

\subsection{Model-free Reinforcement Learning for Eco-driving}
\label{rl_background}
A reinforcement learning environment is commonly formulated as a finite horizon discounted Markov Decision Process (MDP) defined by the six-tuple $\mathcal{M} = \left\langle\mathcal{S},\mathcal{A}, p, r, \rho_{0}, \gamma \right\rangle$ consisting of a state space $\mathcal{S}$, an action space $\mathcal{A}$, a stochastic transition function $p : \mathcal{S} \times \mathcal{A} \rightarrow \Delta(\mathcal{S})$, a reward function $r: \mathcal{S} \times \mathcal{A} \times \mathcal{S} \rightarrow \mathbb{R}$, an initial state distribution $\rho_{0} : \mathcal{S} \rightarrow \Delta(\mathcal{S})$ and a discounting factor $\gamma \in [0,1]$. Here, the probability simplex over $\mathcal{S}$ is defined by the $\Delta (\mathcal{S})$. The objective of RL then becomes searching for a policy $\pi_{\theta} : \mathcal{S} \rightarrow \Delta\mathcal{(A)}$ that maximizes the expected cumulative discounted reward over the MDP. Often a neural network parameterized by $\theta$ will be used as the policy $\pi_{\theta}$. 

\begin{equation}
\max_{\theta} \mathbb{E}_{s_{0} \sim \rho_{0}, a_{t} \sim \pi_{\theta}\left(\cdot | s_{t}\right), s_{t+1} \sim p(\cdot | s_t,a_t)}\left[\sum_{t=0}^{T-1} \gamma^{t} r\left(s_{t}, a_{t}, s_{t+1}\right)\right]
\label{rl-objective}
\end{equation}

In environments where the RL agent can not observe the actual state, an extended MDP formulation containing two more components, observation space $\Omega$ and conditional observation probabilities $O:\mathcal{S}\times\Omega  \rightarrow  \Delta(\Omega)$ can be used. The extended MDP is referred to as a Partially Observable Markov Decision Process (POMDP). In this work, we formulate eco-driving at intersections as a POMDP and use DRL to solve it.




\subsection{Fuel Consumption Model}

In this study we adopt the instantaneous Virginia Tech Comprehensive Power-based Fuel
Consumption Model (VT-CPFM) as our vehicle fuel consumption model~\cite{PARK2013317}. The model provides fairly accurate fuel consumption figures while being easy to calibrate and less computationally expensive. In VT-CPFM, the fuel consumption is modeled as a second order polynomial of vehicle power as given in equation~\ref{fuel-model}. 

\begin{equation}
F(t)=\left\{\begin{array}{cl}\alpha_{0}+\alpha_{1} P(t)+\alpha_{2} P(t)^{2} & \forall P(t) \geq 0 \\ \alpha_{0} & \forall P(t)<0\end{array}\right.
\label{fuel-model}
\end{equation}

Here, instantaneous fuel consumption and vehicle power are denoted by $F(\cdot)$ and $P(\cdot)$ respectively. $\alpha_0, \alpha_1$ and $\alpha_2$ are constants that needs to be calibrated and $t$ is the time step. The instantaneous power $P(\cdot)$ is computed as,

\begin{equation}
 P(t)=\left(\frac{R(t)+1.04 m a(t)}{3600 \eta}\right) v(t)  
\end{equation}

 where the vehicle resistance $R(\cdot)$ is derived from,
 
\begin{equation}
\begin{aligned} R(t)=& \frac{\rho}{25.92} C_{d} C_{a} A_{f} v(t)^{2}+9.8066 m \frac{c_{0}}{1000} c_{1} v(t) \\\\ &+9.8066 m \frac{c_{0}}{1000} c_{2}+9.8066 m G(t) \end{aligned}
\end{equation}

Here, $C_d, C_a$ and $A_f$ are vehicle drag coefficient, altitude correction factor and vehicle front area respectively. $c_0, c_1$ and $c_2$ are rolling resistance related parameters that depends on the road and tire condition. Vehicle mass is given by $m$, instantaneous vehicle acceleration by $a$ and roadway grade by $G$. Air density is denoted by $\rho$ while vehicle driveline efficiency is denoted by $\eta$. In this work, we accommodate the calibrated and validated parameters of VT-CPFM fuel consumption model presented in \cite{PARK2013317, RAKHA2011492}. The parameter configurations for typical gasoline driven passenger cars are shown in Table~\ref{table:fuel-params}.

\begin{table}[]
\centering
\begin{tabular}{ |c|c|c||c|c|c|c| } 
\hline
Parameter & Unit & Value & Parameter & Unit & Value \\
\hline
$\alpha_0$ & - & 0.00078 & m & kg & 3152 \\ 
\hline
$\alpha_1$ & - & 0.000006 & $\eta$ & - & 0.92 \\ 
\hline
$\alpha_2$ & - & 1.9556e-05 & $\rho$ & $kg/m^3$ & 1.23 \\ 
\hline
$c_0$ & - & 1.75 & $C_a$ & - & 0.98\\ 
\hline
$c_1$ & - & 0.033 & $C_d$ & - & 0.6\\ 
\hline
$c_2$ & - & 4.575 & $A_f$ & $m^2$ & 3.28\\ 
\hline
G & - & 0 &  &  & \\ 
\hline
\end{tabular}
\caption{VT-CPFM parameters and their calibrated values}
\label{table:fuel-params}
\end{table}

\subsection{CO$_2$ Emission Model}
Due to the limited availability of efficient emission models, we utilize the default emission model of microscopic traffic simulator SUMO as our CO$_2$ emission model~\cite{hbefa}. SUMO default emission model is based on the Handbook Emission Factors for Road Transport (HBEFA) which includes emission factors for a variety of vehicle categories for different traffic conditions. In this work, we specifically use default HBEFA-v3.1 based CO$_2$ emission model described in~\cite{hbefa} as a third order polynomial. The emission is given as an instantaneous CO$_2$ emission of a vehicle measured in milligrams. In this work, we leverage default SUMO parameter configurations for gasoline driven Euro 4 passenger cars. 

\subsection{Intelligent Driver Model}
\label{idm-discription}
We use the Intelligent Driver Model (IDM) as the car-following model to represent human drivers. As a commonly used car-following model in the research community, IDM can reasonably represent the realistic driver behavior~\cite{Treiber2000CongestedTS} and can produce traffic waves. IDM defined instantaneous acceleration $a(t)$ is given by the Equation~\ref{IDM}, where $v_0, h_0$ and $T$ denote desired velocity, space headway and time headway respectively. Maximum acceleration is given by $c$ and comfortable braking deceleration is given by $b$. Velocity difference between ego vehicle and the leading vehicle is denoted by $\Delta v(t)$ whereas $\delta$ is a constant.

\begin{equation}
a(t)=\frac{d v(t)}{d t}=c\left[1-\left(\frac{v(t)}{v_{0}}\right)^{\delta}-\left(\frac{H(v(t), \Delta v(t)}{h(t)}\right)^{2}\right]
\label{IDM}
\end{equation}

\begin{equation}
H(v(t), \Delta v(t))=h_{0}+\max \left(0, v(t) T+\frac{v(t) \Delta v(t)}{2 \sqrt{c b}}\right)
\end{equation}

The IDM parameters and their calibrated values for human-like driving are given in Table~\ref{table:idm-paramters}.
\begin{table}[]
\centering
\begin{tabular}{ |c|c|c||c|c|c|c| } 
\hline
Parameter & Unit & Value & Parameter & Unit & Value \\
\hline
$v_0$ & m/s & 30 & c & $m/s^2$ & 1 \\ 
\hline
T & s & 1 & b & $m/s^2$ & 1.5 \\ 
\hline
$h_0$ & m & 1.5 & $\delta$ & - & 4 \\ 
\hline
\end{tabular}
\caption{IDM car following model parameters and their calibrated values}
\label{table:idm-paramters}
\end{table}

\section{Method}

In this section, we discuss the DRL based eco-driving controller design for CAVs when approaching and leaving a signalized intersection. We first formally state the problem of eco-driving at signalized intersections and then leverage model-free reinforcement learning in devising the control policy. 

\subsection{Problem Formulation}
\label{optimal-control-problem}

The optimal-control problem for eco-driving at signalized intersections concerns with minimizing the fuel consumption of a fleet of CAVs while having a minimal impact on travel time when the vehicles approach and leave a signalized intersection. Given a vehicle dynamics model $f$ and the traffic signal timing plan $TL_{plan}$ of the immediate signalized intersection, we formulate the optimal-control problem as follows, 

\begin{equation}
\min J= \sum_{i=1}^{n} \int_{0}^{T_i} F\left(a_i(t), v_i(t)\right) dt + T_i
\end{equation}

such that for every vehicle $i$, 

\begin{subequations}
\begin{gather}
a_i(t) = f_i(h_i(t), \dot{h}_i(t), v_i(t)) \\
\int_{0}^{T_i} v_i(t) d t=d \\
h_{min} \leq h_i(t) \leq h_{max} \quad \forall t \in [0, T_i]\\
v_{min} \leq v_i(t) \leq v_{max} \quad \forall t \in [0, T_i]\\
a_{min} \leq a_i(t) \leq a_{max} \quad \forall t \in [0, T_i]
\end{gather}
\end{subequations}

Here $n$ is the number of vehicles, $F(\cdot)$ is the fuel consumption as defined by Equation~\ref{fuel-model}, $T_i$ is the travel time of the CAV $i$ and $d$ is the defined total travel distance. $h(t)$ and $\dot{h}(t)$ denote the headway and relative velocity. $h_{min}$ and $h_{max}$ are the minimum and maximum headway, $v_{min}$ and $v_{max}$ are minimum and maximum velocities, $a_{min}$ and $a_{max}$ are minimum and maximum accelerations, respectively. The three enforced constraints guarantee the safety, connectivity through V2V communication, and passenger comfort.  

\subsection{Model-free Reinforcement Learning for Approximate Control}

The optimal control problem introduced in Section~\ref{optimal-control-problem} is challenging to solve given its non linear nature. A model based approach for solving it usually makes simplifying assumptions about vehicle dynamics, inter-vehicle dynamics apart from requiring significant computation power and thereby being challenging to be used in the real time context. Therefore in this work, we leverage reinforcement learning in solving it approximately. In particular, we formulate the problem of eco-driving at signalized intersections as a discrete time POMDP and use policy gradient methods to solve it. 

We note that our POMDP formulation explained below is defined from a single CAV's perspective. However, we utilize centralized training and a fleet based reward in the POMDP. Therefore, the ultimate control problem we solve is similar to the one presented in Section~\ref{optimal-control-problem}.

\textbf{Assumptions}: We assume each CAV is equipped with on board sensors or V2V (Vehicle to Vehicle) communication with a maximum communication range $r_{v2v}$ for the purpose of communicating with the nearby vehicles. We also assume each CAV can receive Signal Phase and Timing (SPaT) messages from the immediate upcoming traffic signal through I2V (Infrastructure to Vehicle) communications when the vehicle is within the proximity $r_{i2v}$ to the intersection. A schematic overview of these requirements are denoted in Figure~\ref{fig:schematics}.

We characterize the POMDP formulation for Lagrangian control at signalized intersections for energy efficient driving as follows.

\begin{itemize}
    \item 
    \textbf{Observations}: With the mentioned assumptions in place, we define an observation $o \in O$ as an eight tuple ($v_{cav},p_{cav},tl_{phase},v_{lead},p_{lead}, v_{follow},p_{follow}, tl_{time} $) where $v_{cav}, v_{lead}, v_{follow}$ denote the speed of the ego-CAV, immediately leading vehicle and the following vehicle of the ego-CAV. Similarly, $p_{cav}, p_{lead}, p_{follow}$ denote the positions of ego-CAV, lead and follower vehicles. A one hot encoding which gives the phase that the ego-CAV belongs to is denoted by $tl_{phase}$. Finally, $tl_{time}$ is the time to green for the $tl_{phase}$. All observation features are normalized using min-max normalization. 
    Since there are physical limitations on maximum possible communication range in V2V communications, we require $p_{av} - p_{lead}\in (-r_{v2v},r_{v2v})$. Similarly, to ensure uninterrupted I2V communication we require $p_{av} - p_{tl} \in (0,r_{i2v})$ where $p_{tl}$ is the position of the traffic signal.
    
    \item \textbf{Actions}: The action is the continuous acceleration command $a \in (a_{min}, a_{max})$ for the longitudinal control of the CAV.
    
    \item \textbf{Transition Function}: We do not directly define the stochastic transition function of the underlying environment. Instead, advanced microscopic simulation tools are used to sample $s_{t+1} \sim p(s_t, a_t)$. The simulator is in control of applying the accelerations to vehicles and simulates the underlying environment. 
    
    \item \textbf{Reward}: We seek to minimize two objective terms, 1) fuel consumption and 2) impact on travel time. We note that designing a reward function and fine-tuning it is challenging with our two objective terms for several reasons. First, reducing fuel consumption encourage deceleration and thereby low speed. On the other hand, encouraging high average speed reduces the impact on travel time. The two objective terms are therefore competing. Second, the two reward terms are different order polynomials. Therefore, the rate of change of the two reward terms are different in different regions of the composite objective. These two factors make the design of reward function difficult. With manual trial and error, we fine-tuned our reward function to follows. For each phase of the traffic signal, we define the instantaneous reward such that,

\begin{subequations}\label{reward-terms}
\begin{equation}
R_1 = \mu_1 \\
\end{equation}
\begin{equation}
R_2 = \mu_2 + \mu_3\exp{(\Bar{v})} \\
\end{equation}
\begin{equation}
R_3 = \mu_4 + \mu_5\exp{(\Bar{v})} + \mu_6 \Bar{s} \\
\end{equation}
\begin{equation}
R_4 = \mu_{7} + \mu_8 \exp{(\mu_9 \Bar{f})} + \mu_{10}\exp{(\Bar{v})} + \mu_{11} \Bar{s}\\
\end{equation}
\end{subequations}
    
\begin{equation}\label{reward}
  \setlength{\arraycolsep}{0pt}
  r(s,a): = \left\{ \begin{array}{ l l }
    R_1 & \quad \text{if any vehicle stops }\\
    & \quad \textit{at the start of an approach} \\
    R_2 & \quad \text{if} \Bar{f} \leq \delta \wedge \Bar{s} = 0\\
    R_3 & \quad \text{if} \Bar{f} \leq \delta \wedge \Bar{s} > 0\\
    R_4 & \quad \text{otherwise }
  \end{array} \right.
\end{equation}
\end{itemize}

where $\Bar{f}, \Bar{v}, \Bar{s}$ are normalized average fuel consumption per vehicle per phase, normalized average speed per vehicle per phase and normalized number of vehicle stopped on average per phase, respectively. $\delta$ and $\mu_i$ where $i \in [1,\cdots,10]$ are constants and the assigned values are given in Table~\ref{table:reward-paramters}. Both fuel consumption and velocity are normalized based on min-max normalization while per phase number of vehicle stopped are normalized over the total number of vehicle in the phase.  

Reward term $R_1$ in Equation~\ref{reward-terms}a encourages vehicles not to stop at the beginning of an incoming approach. Such a behavior is locally optimal as a stopped vehicle at the start of an incoming approach can block the inflow of vehicles and that leads to low fuel consumption due to lack of vehicles around the intersection. Reward term $R_2$ is defined to encourage high speed of vehicles when the average fuel consumption is below a defined threshold $\delta$. In $R_3$, given that the fuel consumption is still below the defined threshold $\delta$, we penalize low speeds and vehicle stops near the intersection (due to red light) encouraging less stop-and-go behaviors. Finally, in $R_4$, we reward low fuel consumption and high speeds while penalizing vehicle stops to reduce stop-and-go behaviors.

\begin{table}[]
\centering
\begin{tabular}{ |c|c||c|c||c|c|c| } 
\hline
Parameter & Value & Parameter & Value & Parameter & Value \\
\hline
$\mu_1$ & -100 & $\mu_2$  & -5 & $\mu_3$  & 5 \\ 
\hline
$\mu_4$  & -5 & $\mu_5$  & 5 & $\mu_6$  & -10\\ 
\hline
$\mu_7$  & -7 & $\mu_8$  & -3 & $\mu_9$  & 1000\\  
\hline
$\mu_{10}$  & 4 & $\delta$  & 0.01 & $\mu_{11}$  & -10\\  
\hline
\end{tabular}
\caption{Reward function co-coefficients}
\label{table:reward-paramters}
\end{table}

\section{Experimental Setup}

In this section, we discuss the settings used in our numerical experiments. 

\subsection{Network and simulation settings}

SUMO microscopic traffic simulator~\cite{SUMO2018} is used for experimental simulations. A four-way intersection with only through-traffic is designed for the experiments. All incoming and outgoing approaches of the intersection consist of a 250m single lane. For simplicity, we only use standard passenger cars in the experiments. A more realistic traffic mix could be interesting and is kept as future work of this study.  We set speed limits (free-flow speed) of the vehicles to $15 ms^{-1}$. All vehicles enter the system with a speed of $10 ms^{-1}$. For the base experiments, we use 100\% CAV penetration rate and a deterministic vehicle inflow rate of 800 vehicles/hour to simulate vehicle arrivals. 
The traffic signal is designed to carry out a pre-timed fixed time plan with 30 seconds green phase followed by a 4-second yellow phase before turning red. 

\subsection{Training hyperparameters}

Trust Region Policy Optimization (TRPO) algorithm~\cite{pmlr-v37-schulman15} is used to train a neural network with two hidden layers of 64 neurons each and tanh activations as the policy $\pi_{\theta}$. We use centralized training and decentralized execution paradigm with actor critic architecture. A discount factor $\gamma= 0.99$, learning rate $\beta = 0.001$, a training batch size of 55000, and 3000 training iterations are used. We recognize the simulation step duration plays a significant role in minimizing fuel consumption. With a small step duration, CAVs can be controlled more frequently and therefore can achieve better reduction in fuel consumption. However, with small step duration, the training can take significantly more time to converge. To balance the trade-off, we use a simulation step duration of 0.5 seconds with a training horizon of 600 steps in which 100 steps are warm-up steps. Warm-up steps are used to bring the traffic flow to an equilibrium level before introducing control. 

\section{Results}

We evaluate the benefits of proposed Lagrangian control policy for eco-driving against baseline approaches, with the aim of answering the following two questions:

\textbf{Q1} How does the proposed control policy compare to naturalistic driving and model-based control baselines? 

\textbf{Q2} How well does the proposed control policy generalize to environments unseen at training time?

To answer these questions, we develop multiple baseline approaches for eco-driving as follows.

\begin{itemize}
    \item \textbf{V-IDM}: This is the deterministic vanilla version of the IDM car-following model introduced in Section~\ref{idm-discription}. It is a driving baseline that can produce realistic shock waves. 
    \item \textbf{N-IDM}: This baseline adds a noise sampled from a uniform distribution $unif(-0.2,0.2)$ to the IDM defined acceleration $a$. This is a driving baseline that can produce realistic shock waves and models variability in driving behaviors of humans. 
    \item \textbf{M-IDM}: This baseline is developed on top of \textit{N-IDM} model such that IDM parameters introduced in Section~\ref{idm-discription} for each driver are sampled from respective Gaussian distributions. It represents a more diverse mix of drivers with varying levels of aggressiveness.
    \item \textbf{Eco-CACC}: This is a model-based trajectory optimization strategy introduced in~\cite{baseline-one}. We omit the details of the model due to space limitations and refer the reader to \cite{baseline-one} for the implementation details. 
\end{itemize}

\subsection{Analysis of performance}

In answering \textbf{Q1}, we present the performance comparison of the proposed \textit{DRL} control policy vs. the baselines in Table~\ref{compare-q1}. Average per vehicle fuel consumption, emission level and speed are presented. Our proposed control policy is denoted as \textit{DRL}. 

Compared to naturalistic driving, our \textit{DRL} control policy can perform significantly better on all three fronts - fuel, emission, and average speed. This can be observed by comparing the performance against all three \textit{IDM} variants. We specifically note the average speed improvement, which we presume to arise as a result of reducing variability in human driving, alongside the gains from driving energy efficient manner. 

We also note the percentage improvement difference between fuel consumption and emission levels where the emission levels have a greater improvement. We attribute this observation to the model inefficiencies in the SUMO emission model. In particular, the SUMO emission model does not generate any emissions when a vehicle decelerates, leading to an underestimation of the actual emission values. Thus, although our choice for using SUMO emission model was mainly motivated by the computational efficiency and availability, we expect better-calibrated emission models would produce similar improvements as in fuel consumption improvements.

In comparison to \textit{Eco-CACC}, our \textit{DRL} control policy performs better in fuel consumption and emission levels while maintaining a similar average speed per vehicle. We note that our \textit{DRL} policy achieves slightly better performance than \textit{Eco-CACC} without having access to information such as queue discharge rates at intersections and without assuming a vehicle dynamics model as \textit{Eco-CACC} does. 

\begin{figure*}[bt!]
\centering
\begin{subfigure}{0.31\linewidth}
  \centering
  \includegraphics[width=\linewidth]{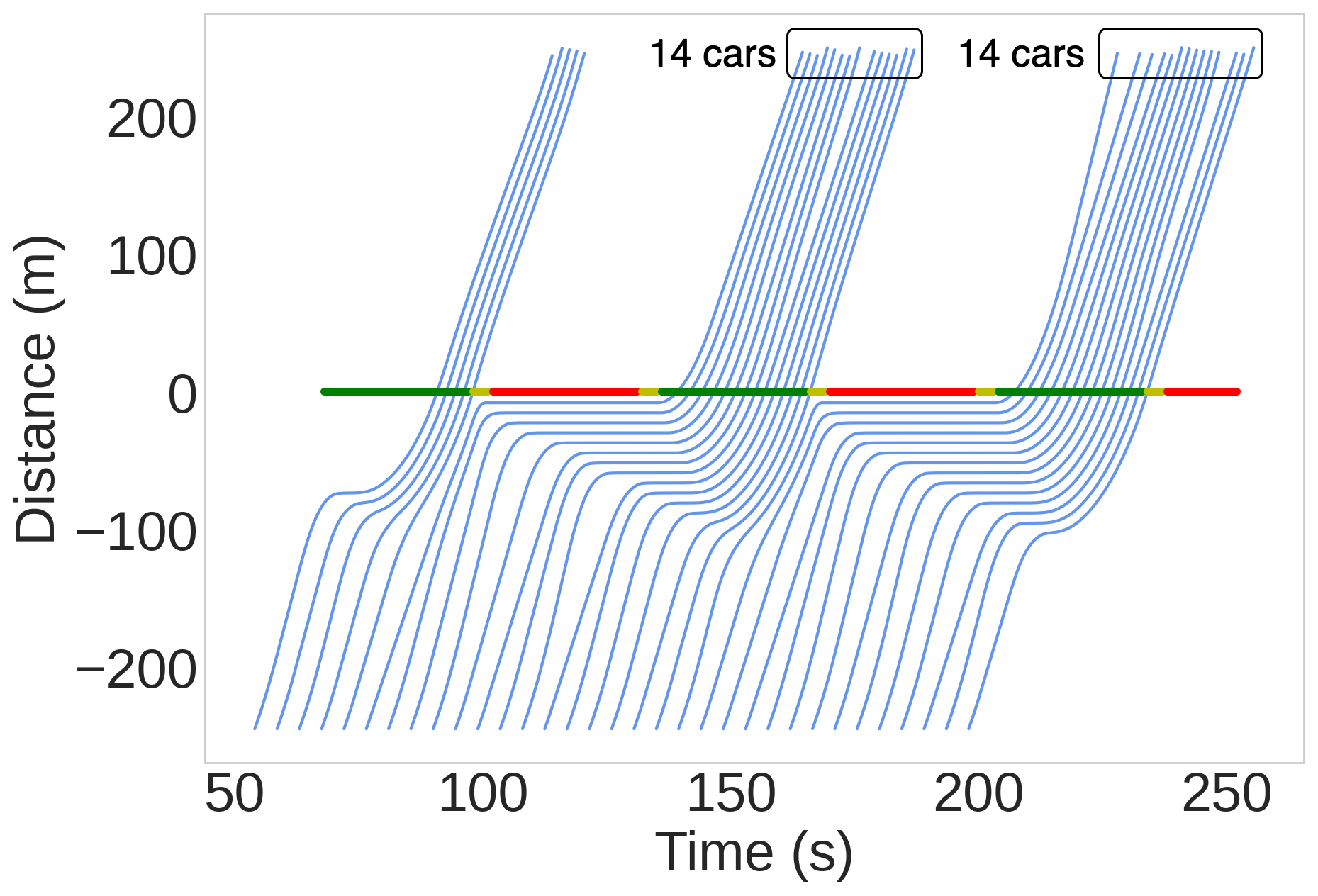}
  \caption{Human driven vehicles based on V-IDM model}
  \label{fig11a}
\end{subfigure}
\begin{subfigure}{0.31\linewidth}
  \centering
  \includegraphics[width=\linewidth]{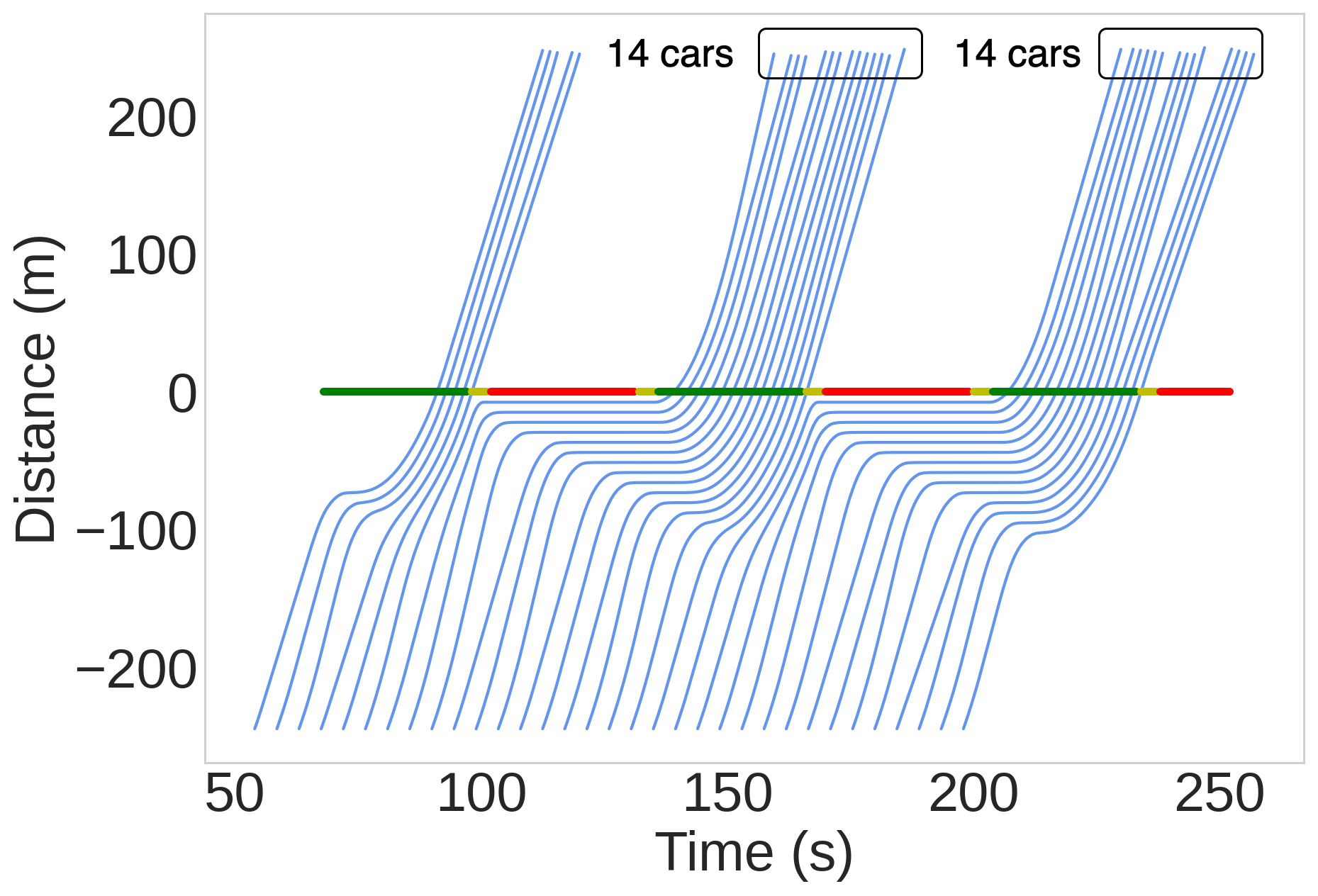}
  \caption{Human driven vehicles based on N-IDM model}
  \label{fig11b}
\end{subfigure}
\begin{subfigure}{0.31\linewidth}
  \centering
  \includegraphics[width=\linewidth]{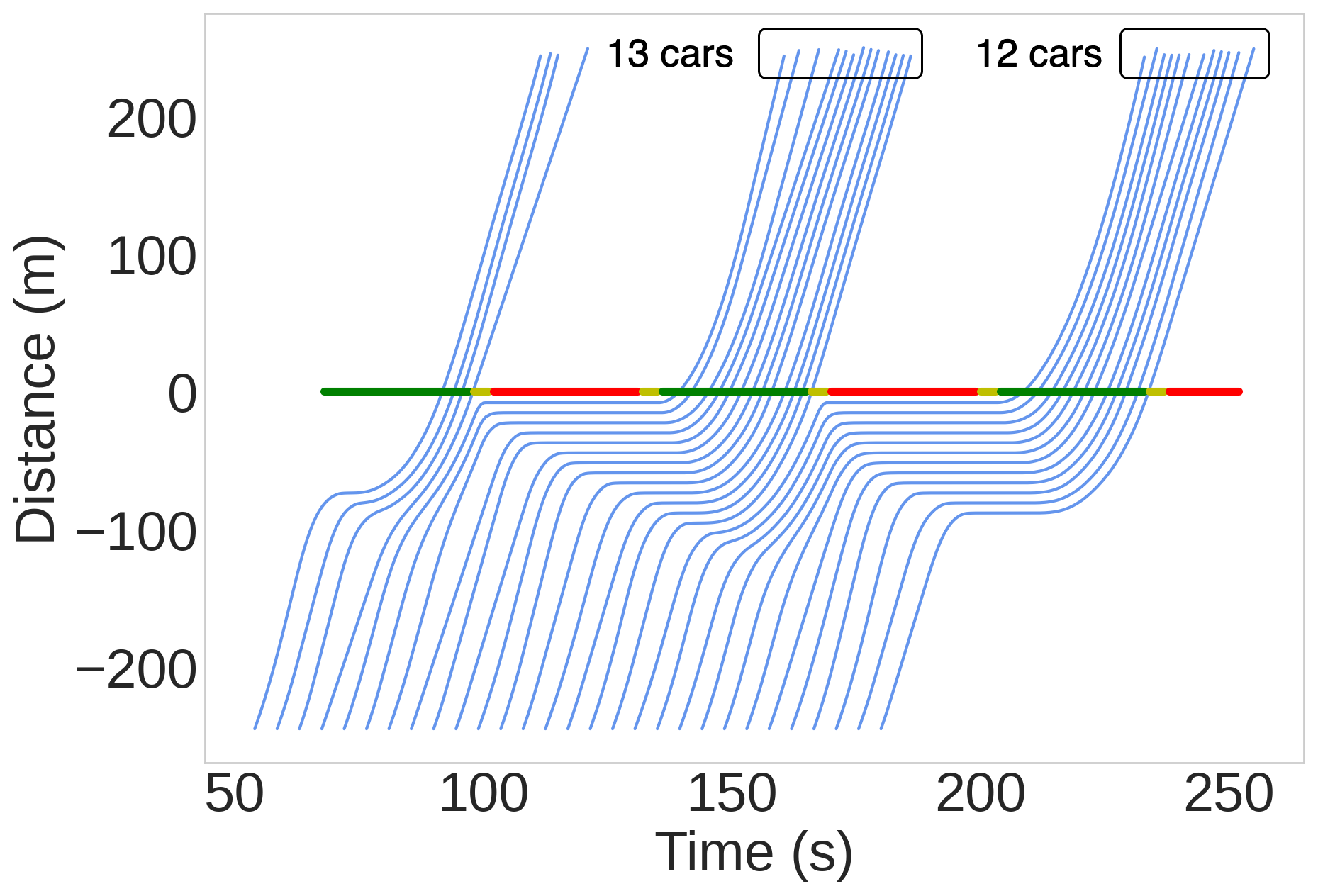}
  \caption{Human driven vehicles based on M-IDM model}
  \label{fig11b}
\end{subfigure}
\begin{subfigure}{0.31\linewidth}
  \centering
  \includegraphics[width=\linewidth]{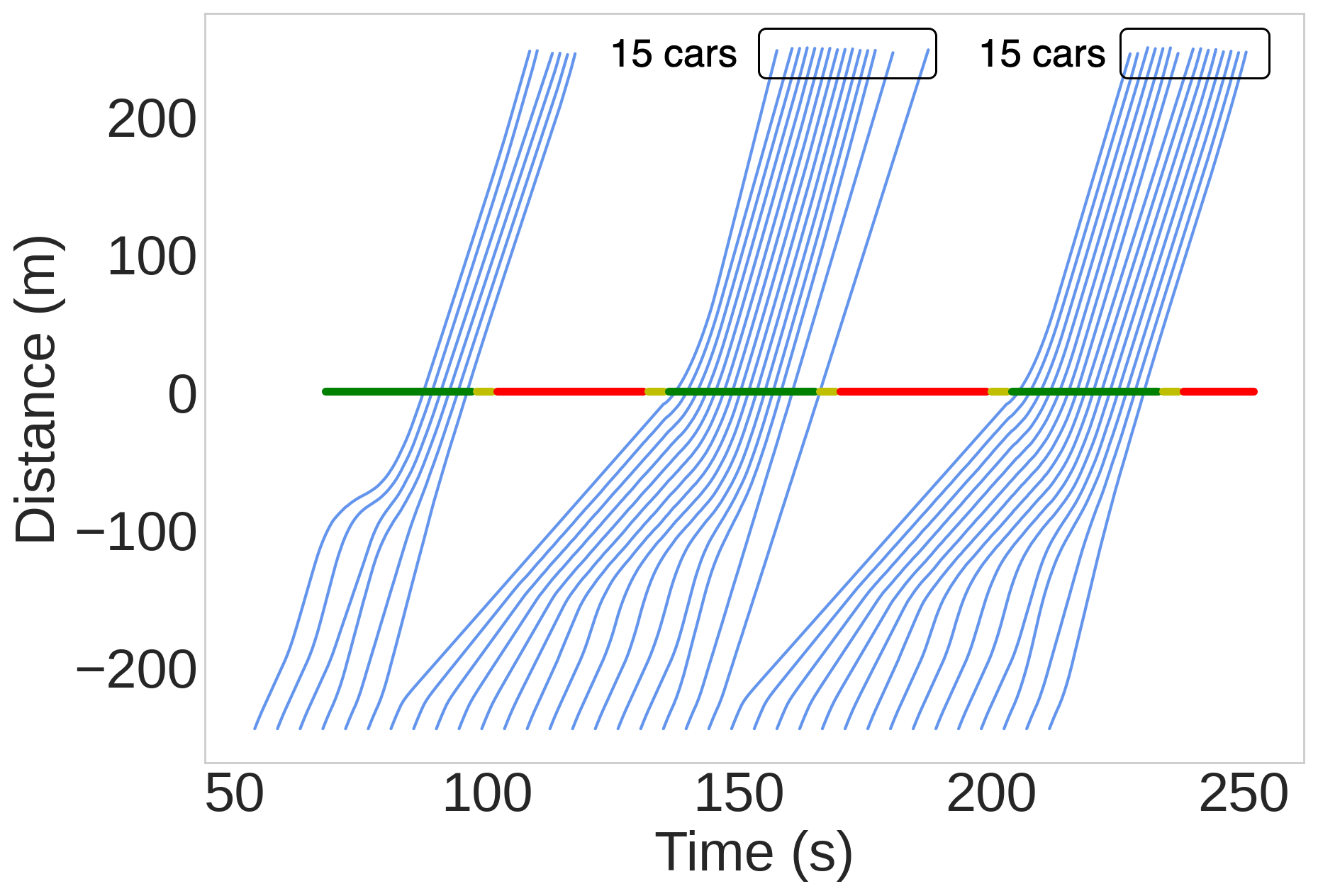}
  \caption{Eco-CACC vehicles }
  \label{fig11a}
\end{subfigure}
\begin{subfigure}{0.31\linewidth}
  \centering
  \includegraphics[width=\linewidth]{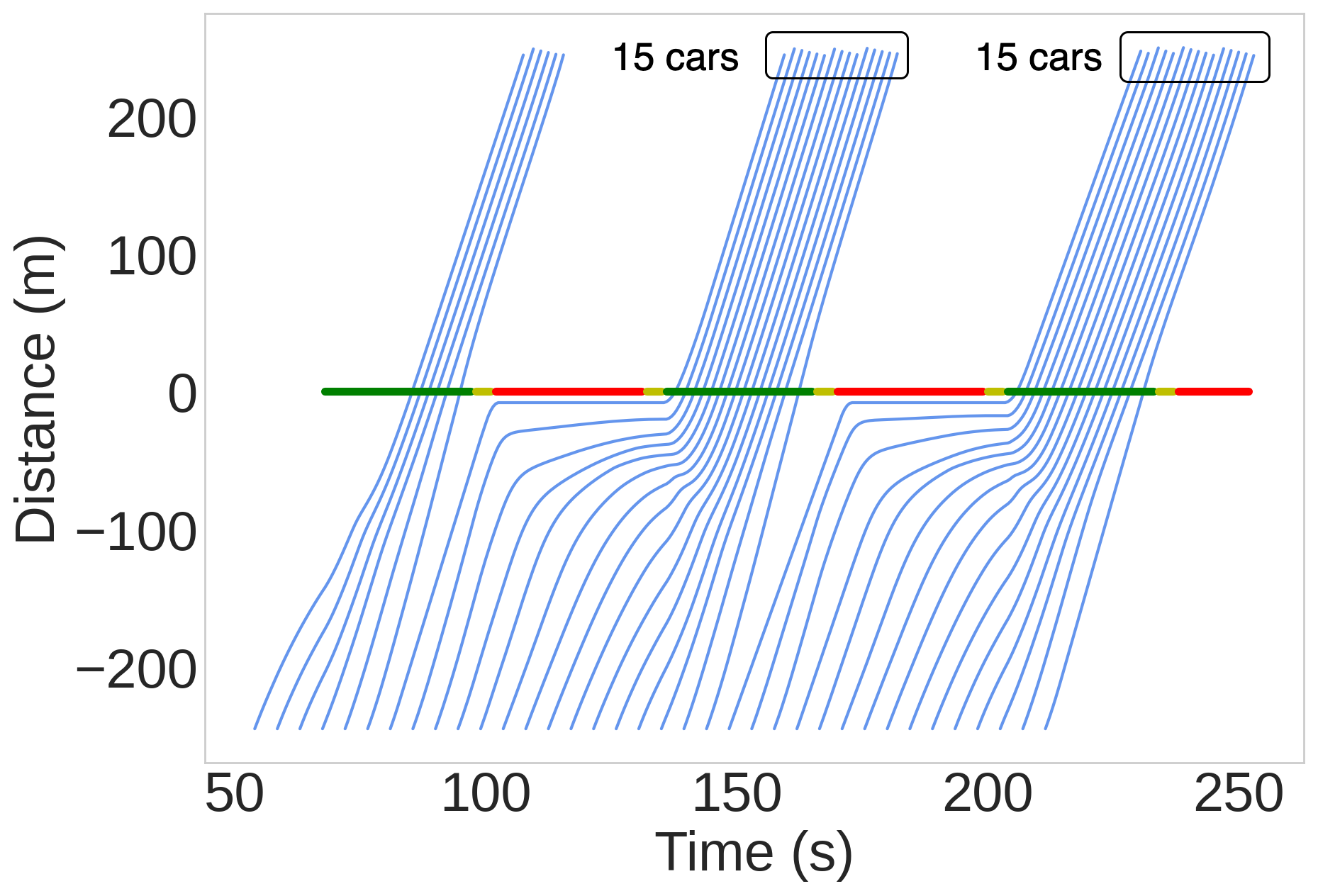}
  \caption{DRL model based vehicles}
  \label{fig11b}
\end{subfigure}
\begin{subfigure}{0.29\linewidth}
  \centering
  \includegraphics[width=\linewidth]{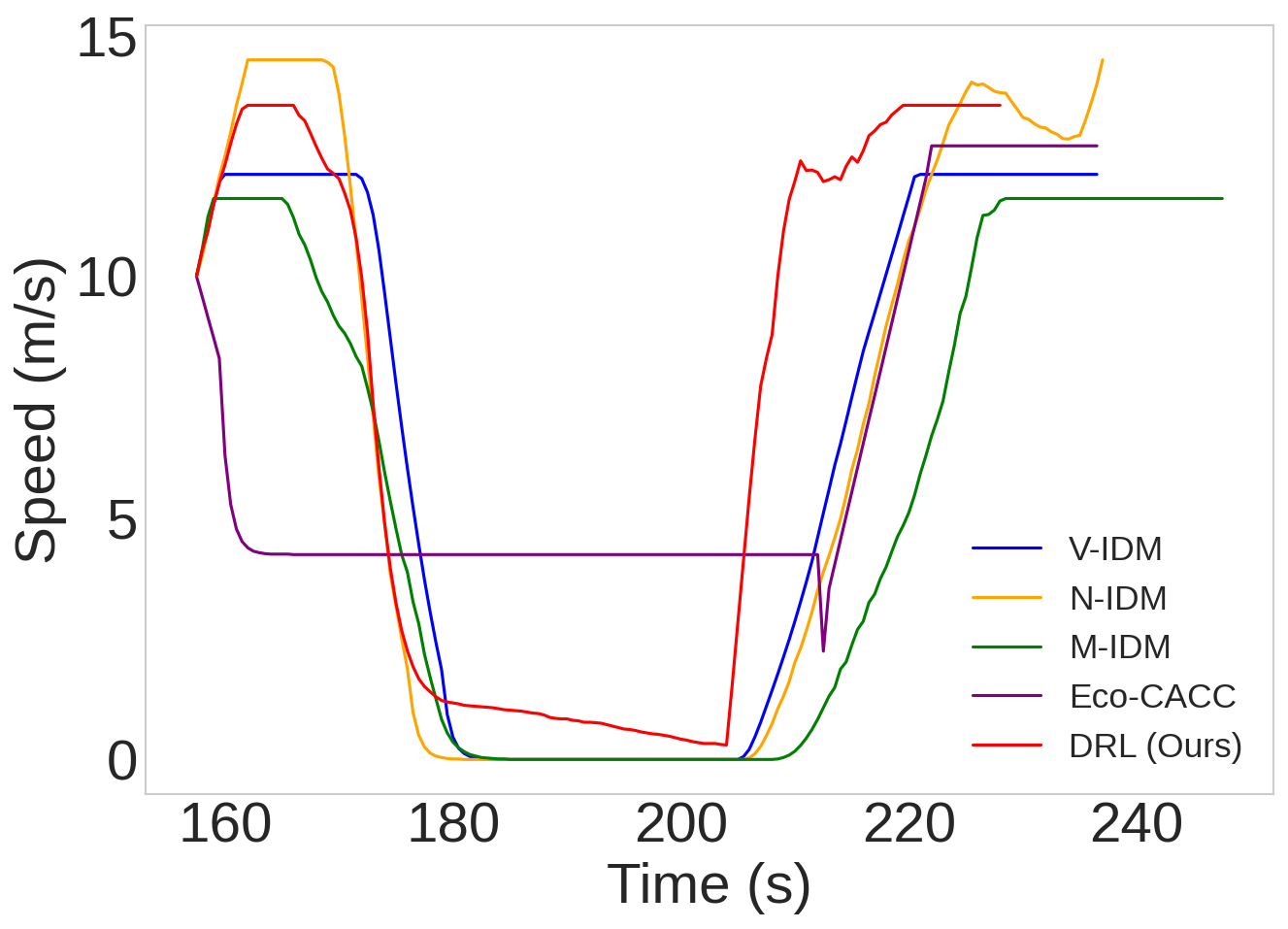}
  \caption{Speed profiles}
  \label{fig11b}
\end{subfigure}
\caption{Time space diagrams of north-bound vehicle trajectories produced by a) V-IDM model, b) N-IDM model, c) M-IDM model, d) Eco-CACC model, and e) DRL model. Both Eco-CACC and DRL models demonstrate behaviors which involve reduced stopping at the intersection as compared to IDM variants. Both Eco-CACC and DRL models increase the throughput of vehicles during a green light phase by one extra vehicle as can be seen in Figure~\ref{ts-diagrams}d and \ref{ts-diagrams}e. Figure~\ref{ts-diagrams}f shows the speed profile of a selected vehicle under the five different control strategies.}
\label{ts-diagrams}
\end{figure*}

To further investigate the behaviors of these policies, we present the time-space diagrams of each policy behavior in Figure~\ref{ts-diagrams}. As can be seen from Figure~\ref{ts-diagrams}a, \ref{ts-diagrams}b and \ref{ts-diagrams}c, all IDM-based variants stop the vehicles at the intersection creating stop-and-go waves. Also, with imperfect driving and varying aggressiveness levels, the number of vehicles that can cross the intersection within a green phase decreases, as shown by the annotations on each of the figures. Under 100\% CAV penetration rate, \textit{Eco-CACC} model does not create any stop-and-go waves as shown in Figure~\ref{ts-diagrams}d. Similarly, our \textit{DRL} model demonstrates a not stopping behavior for the majority of the vehicles while only one vehicle stops at the intersection. However, 15 vehicles are crossing the intersection under both \textit{DRL} and \textit{Eco-CACC} models within a single green phase, giving a throughput advantage over all IDM variants. 

We hypothesize that the reason for one vehicle stopping at the intersection under \textit{DRL} model is aligned with the composite reward function we utilized and how the state features are used by the policy to control the CAV. In particular, we found that there are two types of behaviors that the \textit{DRL} based control policy has to learn when controlling CAVs that are approaching the intersection: 1) the behavior of the leading vehicle (CAV that is closest to the intersection) and 2) the behavior of the following vehicles. The leading vehicle must learn to utilize the time remaining in the current phase to decide an acceleration to achieve the not stopping behavior. However, all the other following vehicles only need to observe their immediate leading vehicle to achieve not stopping behavior and utilize the remaining phase time only to calibrate their action. This creates two different mappings from states to actions. We expect further reward tuning and state-space refinements or training separate policies for leading and non-leading CAVs could result in the leading vehicle demonstrating the not stopping behavior, further reducing fuel consumption and emission levels. 

In Figure~\ref{ts-diagrams}f, we present the speed profile of the same vehicle under the five different control strategies. Interestingly, \textit{DRL} policy-based speed profile takes a profile that is closer to naturalistic driving than \textit{Eco-CACC}. This may actually be a desired behavior of a CAV control policy, especially in a context like mixed traffic where there is a mix of human drivers and CAVs. In such cases, it is desirable to have a CAV control policy that is more familiar to the human drivers who are following such a CAV. Often referred in the literature as socially compatible driving design~\cite{socially-compatible}, CAVs are expected to consider the impacts of their actions on the following human drivers in car-following scenarios. Therefore, the velocity profile in Figure~\ref{ts-diagrams}f under \textit{DRL} policy can be treated as socially compatible design due to its similarity to human-like driving behaviors (\textit{V-IDM}, \textit{N-IDM} and \textit{M-IDM}).


\begin{table}[]
\centering
\begin{tabular}{lccc}
\hline \multirow{1}{0pt}{Model} & \multicolumn{3}{c}{Metric} \\
\cline {2 - 4} & Fuel(L) & Emission(kg) & Avg. speed(m/s) \cr
\hline
V-IDM & 0.1160 & 0.1308 & 3.96 \cr
N-IDM & 0.1169 & 0.1345 & 3.94 \cr 
M-IDM & 0.1281 & 0.1465 & 3.60 \cr 
Eco-CACC & 0.1001 & 0.1006 & $\mathbf{4.80}$ \cr 
DRL & $\mathbf{0.0954}$ & $\mathbf{0.0976}$ & ${4.75}$ \cr 
\hline
Gain (vs V-IDM) & ${17.76\%}$ & ${25.38\%}$ & ${19.95\%}$ \cr 
Gain (vs N-IDM) & ${18.39\%}$ & ${27.43\%}$ & ${20.56\%}$ \cr 
Gain (vs M-IDM) & ${25.53\%}$ & ${33.38\%}$ & ${31.94\%}$ \cr 
Gain (vs Eco-CACC) & ${4.70\%}$ & ${2.98\%}$ & ${-1.04\%}$ \cr 
\hline
\end{tabular}
\caption{Comparison of per vehicle fuel consumption (lower is better), emission level (lower is better) and average speed (higher is better) under different control strategies with 100\% CAV penetration rate.}
\label{compare-q1}
\end{table}

\subsection{Generalizability of the control policy}

\begin{figure*}[bt!]
\centering
\begin{subfigure}{0.31\linewidth}
  \centering
  \includegraphics[width=\linewidth]{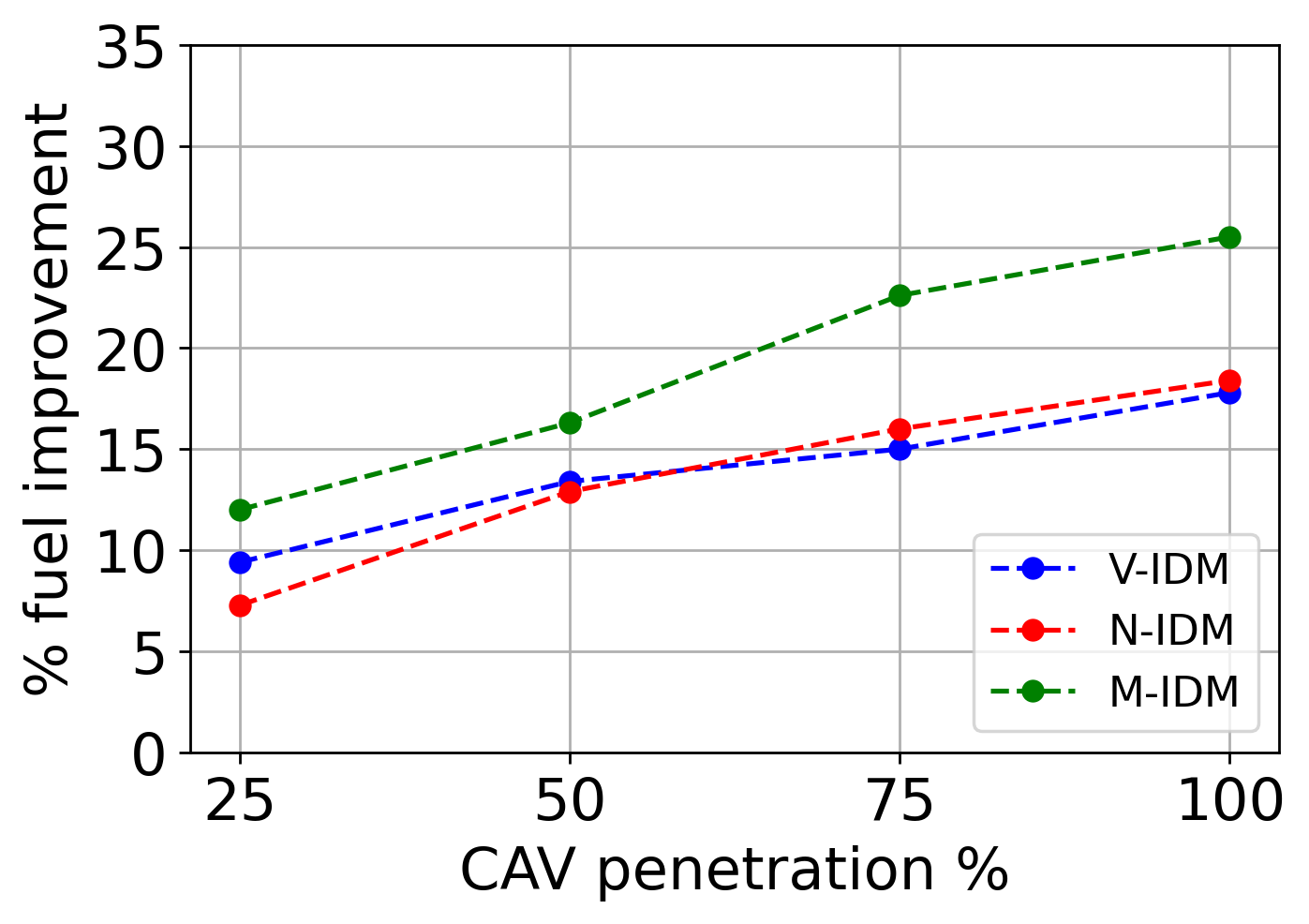}
  \caption{Percentage fuel usage improvement}
  \label{fig11a}
\end{subfigure}
\begin{subfigure}{0.31\linewidth}
  \centering
  \includegraphics[width=\linewidth]{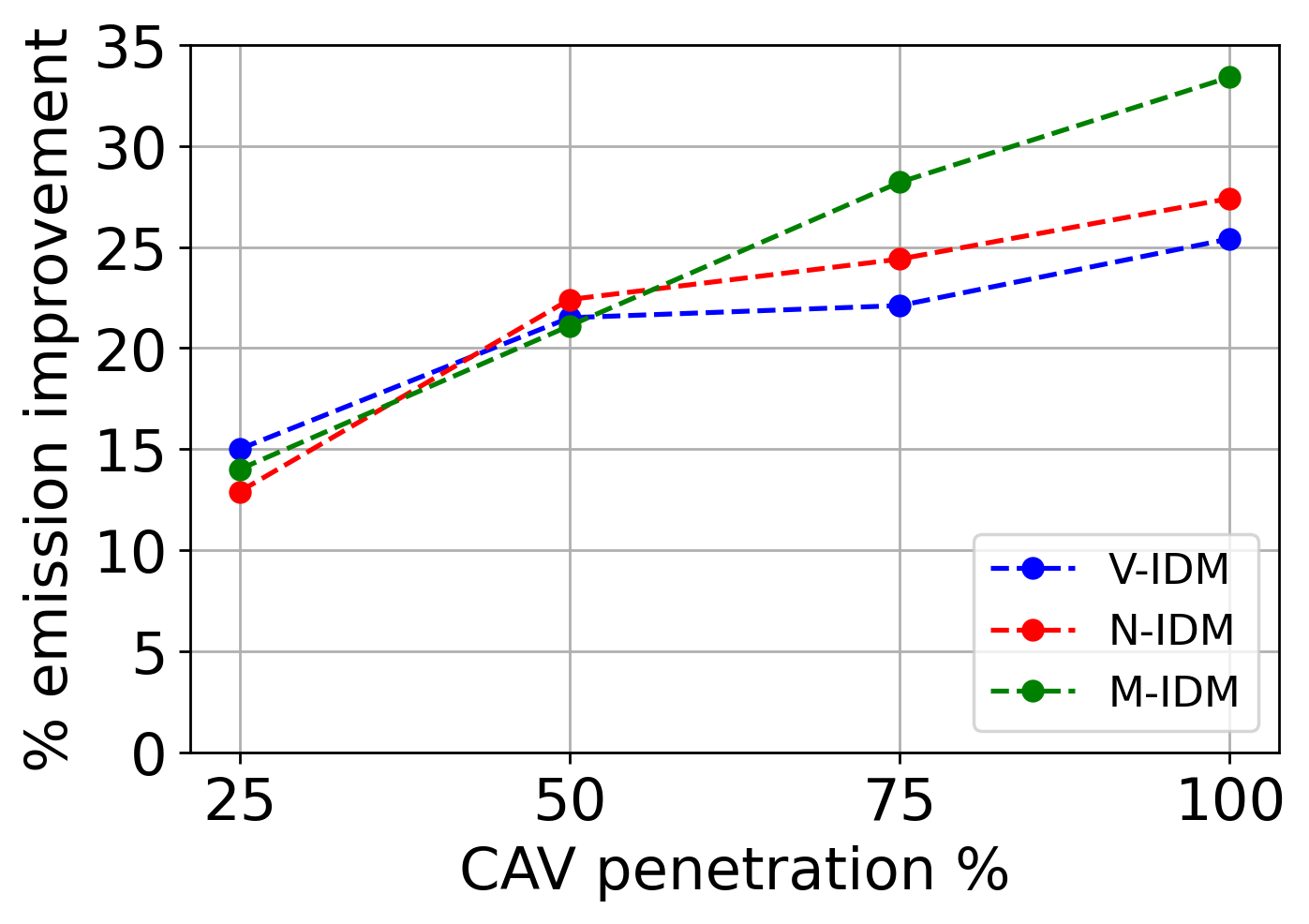}
  \caption{Percentage emission level improvement}
  \label{fig11b}
\end{subfigure}
\begin{subfigure}{0.31\linewidth}
  \centering
  \includegraphics[width=\linewidth]{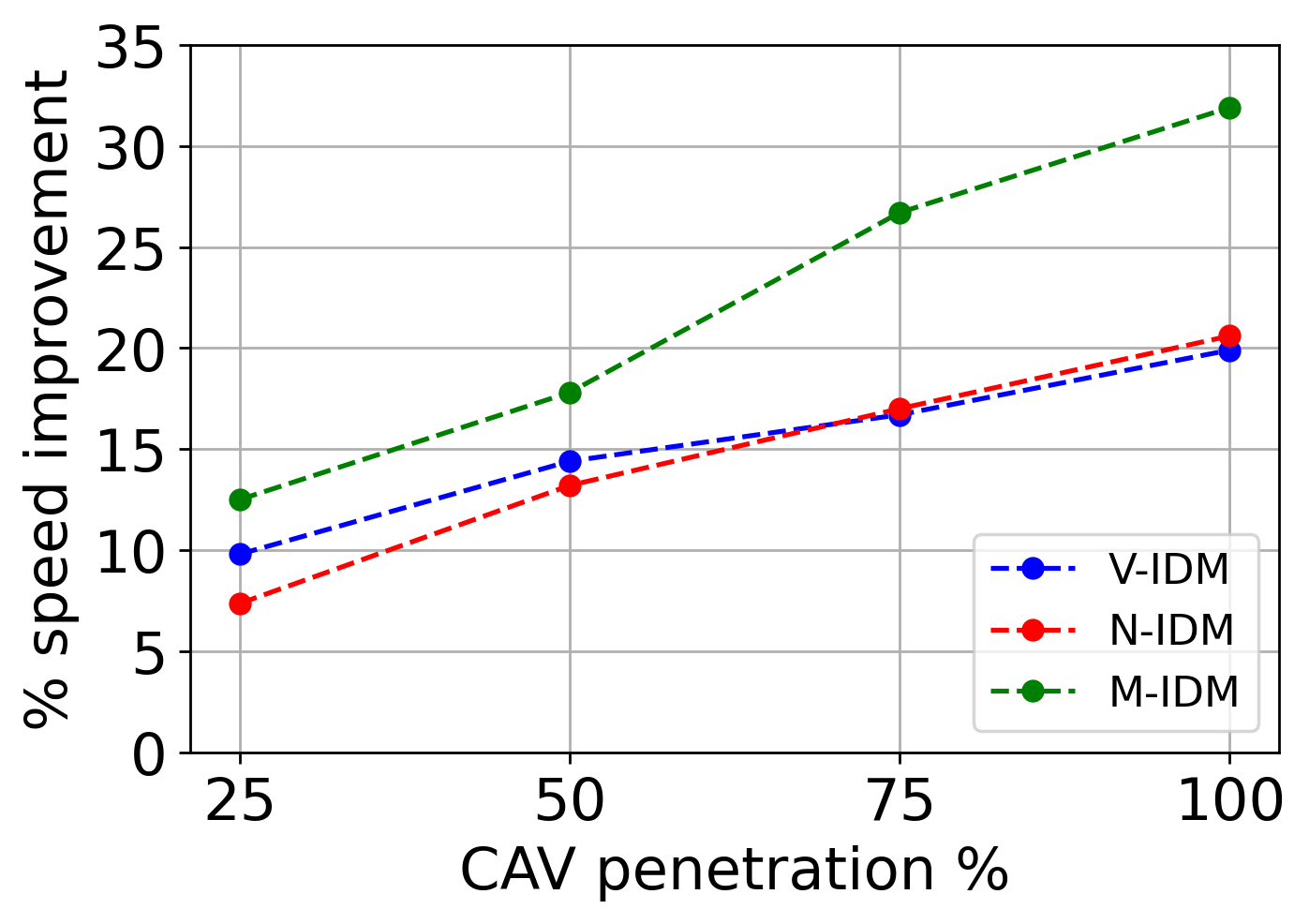}
  \caption{Percentage speed improvement}
  \label{fig11b}
\end{subfigure}
\caption{Percentage improvement in terms of fuel usage, emission levels and average speed from the IDM variant baselines given in Table IV under different CAV penetration rates (CAVs are controlled by zero-shot transferred DRL policy).}
\label{cav-zero-diagrams}
\end{figure*}

In answering \textbf{Q2}, we analyze the zero-shot transfer capacity of the learned policy to unseen mixed traffic scenarios. The results are shown in Table~\ref{compare-q2} and in Figure~\ref{cav-zero-diagrams}. In these experiments, a fraction of CAVs is controlled by the \textit{DRL} policy directly applied from 100\% CAVs based training (zero-shot transfer). The remaining vehicles are human-driven vehicles (and thus create the context of mixed traffic) controlled by one of the three naturalistic driving baselines: \textit{V-IDM, N-IDM, M-IDM}. Since in the 100\% CAVs based training, the policy has not seen human-like naturalistic driving behaviors, the mixed traffic scenarios naturally fall under the context of out-of-distribution.

As can be seen, Table~\ref{compare-q2} and Figure~\ref{cav-zero-diagrams} indicate promising results in terms of the generalizability of the learned policy. As expected, the 25\% CAV penetration level produces the lowest performance while 100\% produces the highest. This phenomenon can be intuitively understood given that more CAVs can introduce more control to the system. It should be noted that even with 25\% CAV penetration rate, the \textit{DRL} policy can still achieve significant benefits on all three fronts (fuel, emission and average speed). In general, for both fuel and emission, 25\% CAVs bring at least 40\% of the total benefit where it is more than 50\% when the \textit{V-IDM} model is used for human driving. Furthermore, the out-of distribution performance of the learned policy under the noisy and varying aggressiveness-based IDM variants demonstrates a highly favorable trait of a desired eco-driving strategy--the possibility to perform under unexpected events--which is challenging to achieve with model-based methods.    

\begin{table}[]
\centering
\begin{tabular}{lccc}
\hline \multirow{1}{0pt}{CAV \%} & \multicolumn{3}{c}{Percentage improvement from baseline (Table~\ref{compare-q1})} \\
\cline {2 - 4} & V-IDM (F$\vert$E$\vert$S)\% & N-IDM (F$\vert$E$\vert$S)\% & M-IDM (F$\vert$E$\vert$S)\% \cr
\hline
25 & 9.40$\vert$15.0$\vert$9.80 & 7.27$\vert$12.9$\vert$7.36 &12.0$\vert$14.0$\vert$12.5 \cr
50 & 13.4$\vert$21.5$\vert$14.4 & 12.9$\vert$22.4$\vert$13.2 &16.3$\vert$21.1$\vert$17.8 \cr
75 & 15.0$\vert$22.1$\vert$16.7 & 16.0$\vert$24.4$\vert$17.0 &22.6$\vert$28.2$\vert$26.7 \cr 
100 & 17.8$\vert$25.4$\vert$19.9 & 18.4$\vert$27.4$\vert$20.6 &25.5$\vert$33.4$\vert$31.9 \cr 
\hline
\end{tabular}
\caption{Percentage improvement from the IDM variant baselines given in Table~\ref{compare-q1} in terms of per vehicle fuel consumption, emission level and average speed under different CAV penetration rates controlled by zero-shot transferred \textit{DRL} policy. Here F denotes fuel consumption percentage improvement, E denotes emission level percentage improvement and S denotes average speed percentage improvement.}.
\label{compare-q2}
\end{table}

\section{Conclusion and Future work}

In this work, we study eco-driving at signalized intersections in which we seek to reduce fuel consumption and emission levels while having a minimal impact on the travel time of the vehicles. 
Future work of this study can span out in multiple directions. First, one interesting direction is to consider a fleet CAVs under multiple signalized intersections to quantitatively assess the benefit of reinforcement learning for eco-driving at intersections. Second, as we have found, designing a composite reward function with multiple competing objectives is challenging. Given that the ultimate gain of the proposed method significantly depends on the design of the reward function, further studies that can shed light in this direction are encouraged. 

\section{Acknowledgement}

The authors acknowledge the MIT SuperCloud and Lincoln Laboratory Supercomputing Center for providing computational resources that have contributed to the research results reported within this paper. The authors are grateful to Mark Taylor, Blaine Leonard, Matt Luker and Michael Sheffield at the Utah Department of Transportation for numerous constructive discussions. The authors would also like to thank Zhongxia Yan for his help in developing the original framework that was extended to produce the computational results reported in this paper.

\bibliographystyle{unsrt} 
\bibliography{references.bib}

\begin{thebibliography}{10}

\bibitem{energy-literacy}
Energy literacy.
\newblock \url{http://energyliteracy.com}.
\newblock [Online; accessed 22-04-2022].

\bibitem{emission-waves}
Hesham Rakha, Kyoungho Ahn, and Antonio Trani.
\newblock Comparison of mobile5a, mobile6, vt-micro, and cmem models for
  estimating hot-stabilized light-duty gasoline vehicle emissions.
\newblock {\em Canadian Journal of Civil Engineering}, 30(6):1010--1021, 2003.

\bibitem{real-world-emission}
Matthew Barth and Kanok Boriboonsomsin.
\newblock Real-world carbon dioxide impacts of traffic congestion.
\newblock {\em Transportation Research Record}, 2008.

\bibitem{article}
Yonglian Ding.
\newblock Trip-based explanatory variables for estimating vehicle fuel
  consumption and emission rates.
\newblock {\em Water Air and Soil Pollution Focus}, 2:61--77, 09 2002.

\bibitem{Yang2021EcoDrivingAS}
Hao Yang, Fawaz Almutairi, and Hesham~A. Rakha.
\newblock Eco-driving at signalized intersections: A multiple signal
  optimization approach.
\newblock {\em IEEE Transactions on Intelligent Transportation Systems}, 2021.

\bibitem{Cui2018ImpactOA}
Lian Cui, Huifu Jiang, B.~Brian Park, Young-Ji Byon, and Jia Hu.
\newblock Impact of automated vehicle eco-approach on human-driven vehicles.
\newblock {\em IEEE Access}, 6:62128--62135, 2018.

\bibitem{Ozkan2020APC}
M.~Ozkan and Yao Ma.
\newblock A predictive control design with speed previewing information for
  vehicle fuel efficiency improvement.
\newblock {\em 2020 American Control Conference (ACC)}, pages 2312--2317, 2020.

\bibitem{wu2017emergent}
Cathy Wu, Aboudy Kreidieh, Eugene Vinitsky, and Alexandre~M Bayen.
\newblock Emergent behaviors in mixed-autonomy traffic.
\newblock In {\em Conference on Robot Learning}, 2017.

\bibitem{flow}
Cathy Wu, Abdul~Rahman Kreidieh, Kanaad Parvate, Eugene Vinitsky, and
  Alexandre~M Bayen.
\newblock Flow: A modular learning framework for mixed autonomy traffic.
\newblock {\em IEEE Transactions on Robotics}, 2021.

\bibitem{Yu2019ReinforcementLI}
Chao Yu, Jiming Liu, and Shamim Nemati.
\newblock Reinforcement learning in healthcare: A survey.
\newblock 2019.

\bibitem{RL-finance}
Arthur Charpentier, Romuald Elie, and Carl Remlinger.
\newblock {Reinforcement Learning in Economics and Finance}.
\newblock Technical report, 2020.

\bibitem{FONTARAS201797}
Georgios Fontaras, Nikiforos-Georgios Zacharof, and Biagio Ciuffo.
\newblock Fuel consumption and co2 emissions from passenger cars in europe –
  laboratory versus real-world emissions.
\newblock {\em Progress in Energy and Combustion Science}, 60:97--131, 2017.

\bibitem{eco-platoon}
Ziran Wang, Guoyuan Wu, Peng Hao, Kanok Boriboonsomsin, and Matthew Barth.
\newblock Developing a platoon-wide eco-cooperative adaptive cruise control
  (cacc) system.
\newblock In {\em Intelligent Vehicles Symposium}, 2017.

\bibitem{BARTH2009400}
Matthew Barth and Kanok Boriboonsomsin.
\newblock Energy and emissions impacts of a freeway-based dynamic eco-driving
  system.
\newblock {\em Transportation Research Part D: Transport and Environment},
  14(6):400--410, 2009.
\newblock The interaction of environmental and traffic safety policies.

\bibitem{baseline-one}
Hao Yang, Hesham Rakha, and Mani~Venkat Ala.
\newblock Eco-cooperative adaptive cruise control at signalized intersections
  considering queue effects.
\newblock {\em IEEE Transactions on Intelligent Transportation Systems},
  18(6):1575--1585, 2017.

\bibitem{ZHAO2018802}
Weiming Zhao, Dong Ngoduy, Simon Shepherd, Ronghui Liu, and Markos
  Papageorgiou.
\newblock A platoon based cooperative eco-driving model for mixed automated and
  human-driven vehicles at a signalised intersection.
\newblock {\em Transportation Research Part C: Emerging Technologies}.

\bibitem{mpc-1}
Baisravan HomChaudhuri, Ardalan Vahidi, and Pierluigi Pisu.
\newblock A fuel economic model predictive control strategy for a group of
  connected vehicles in urban roads.
\newblock In {\em 2015 American Control Conference (ACC)}, pages 2741--2746,
  2015.

\bibitem{dp-1}
Chao Sun, Xinwei Shen, and Scott Moura.
\newblock Robust optimal eco-driving control with uncertain traffic signal
  timing.
\newblock In {\em 2018 Annual American Control Conference (ACC)}, pages
  5548--5553, 2018.

\bibitem{BEUSEN2009514}
Bart Beusen, Steven Broekx, Tobias Denys, Carolien Beckx, Bart Degraeuwe,
  Maarten Gijsbers, Kristof Scheepers, Leen Govaerts, Rudi Torfs, and Luc~Int
  Panis.
\newblock Using on-board logging devices to study the longer-term impact of an
  eco-driving course.
\newblock {\em Transportation Research Part D: Transport and Environment},
  14(7):514--520, 2009.

\bibitem{ZARKADOULA2007449}
Maria Zarkadoula, Grigoris Zoidis, and Efthymia Tritopoulou.
\newblock Training urban bus drivers to promote smart driving: A note on a
  greek eco-driving pilot program.
\newblock {\em Transportation Research Part D: Transport and Environment},
  12(6):449--451, 2007.

\bibitem{gov-eco-test}
Jiaqi Ma, Jia Hu, Ed~Leslie, Fang Zhou, and Zhitong Huang.
\newblock Eco-drive experiment on rolling terrain for fuel consumption
  optimization – summary report.
\newblock {\em Transport Research International Documentation}.

\bibitem{8917077}
Ziran Wang, Yuan-Pu Hsu, Alexander Vu, Francisco Caballero, Peng Hao, Guoyuan
  Wu, Kanok Boriboonsomsin, Matthew~J. Barth, Aravind Kailas, Pascal Amar,
  Eddie Garmon, and Sandeep Tanugula.
\newblock Early findings from field trials of heavy-duty truck connected
  eco-driving system.
\newblock In {\em 2019 IEEE Intelligent Transportation Systems Conference
  (ITSC)}, pages 3037--3042, 2019.

\bibitem{bottleneck}
Eugene Vinitsky, Kanaad Parvate, Aboudy Kreidieh, Cathy Wu, and Alexandre
  Bayen.
\newblock Lagrangian control through deep-rl: Applications to bottleneck
  decongestion.
\newblock In {\em 2018 21st International Conference on Intelligent
  Transportation Systems (ITSC)}, pages 759--765, 2018.

\bibitem{Yan_2021}
Zhongxia Yan and Cathy Wu.
\newblock Reinforcement learning for mixed autonomy intersections.
\newblock In {\em 2021 {IEEE} International Intelligent Transportation Systems
  Conference ({ITSC})}, 2021.

\bibitem{intellight}
Hua Wei, Guanjie Zheng, Huaxiu Yao, and Zhenhui Li.
\newblock Intellilight: A reinforcement learning approach for intelligent
  traffic light control.
\newblock KDD '18, page 2496–2505, 2018.

\bibitem{DTlight}
Vindula Jayawardana, Anna Landler, and Cathy Wu.
\newblock Mixed autonomous supervision in traffic signal control.
\newblock In {\em 2021 IEEE International Intelligent Transportation Systems
  Conference (ITSC)}, 2021.

\bibitem{Zhu2021ADR}
Zhaoxuan Zhu, Shobhit Gupta, Abhishek Gupta, and Marcello Canova.
\newblock A deep reinforcement learning framework for eco-driving in connected
  and automated hybrid electric vehicles.
\newblock 2021.

\bibitem{GUO2021102980}
Qiangqiang Guo, Ohay Angah, Zhijun Liu, and Xuegang~(Jeff) Ban.
\newblock Hybrid deep reinforcement learning based eco-driving for low-level
  connected and automated vehicles along signalized corridors.
\newblock {\em Transportation Research Part C: Emerging Technologies},
  124:102980, 2021.

\bibitem{WEGENER2021102967}
Marius Wegener, Lucas Koch, Markus Eisenbarth, and Jakob Andert.
\newblock Automated eco-driving in urban scenarios using deep reinforcement
  learning.
\newblock {\em Transportation Research Part C: Emerging Technologies},
  126:102967, 2021.

\bibitem{PARK2013317}
Sangjun Park, Hesham Rakha, Kyoungho Ahn, and Kevin Moran.
\newblock Virginia tech comprehensive power-based fuel consumption model
  (vt-cpfm): Model validation and calibration considerations.
\newblock {\em International Journal of Transportation Science and Technology},
  2013.

\bibitem{RAKHA2011492}
Virginia tech comprehensive power-based fuel consumption model: Model
  development and testing.
\newblock {\em Transportation Research Part D: Transport and Environment},
  16(7):492--503, 2011.

\bibitem{hbefa}
Daniel Krajzewicz, Michael Behrisch, Peter Wagner, Raphael Luz, and Mario
  Krumnow.
\newblock Second generation of pollutant emission models for sumo.
\newblock In Michael Behrisch and Melanie Weber, editors, {\em Modeling
  Mobility with Open Data}, pages 203--221, Cham, 2015. Springer International
  Publishing.

\bibitem{Treiber2000CongestedTS}
Treiber, Hennecke, and Helbing.
\newblock Congested traffic states in empirical observations and microscopic
  simulations.
\newblock {\em Physical review. E}, 2000.

\bibitem{SUMO2018}
Pablo~Alvarez Lopez, Michael Behrisch, Laura Bieker-Walz, Jakob Erdmann,
  Yun-Pang Fl{\"o}tter{\"o}d, Robert Hilbrich, Leonhard L{\"u}cken, Johannes
  Rummel, Peter Wagner, and Evamarie Wie{\ss}ner.
\newblock Microscopic traffic simulation using sumo.
\newblock 2018.

\bibitem{pmlr-v37-schulman15}
John Schulman, Sergey Levine, Pieter Abbeel, Michael Jordan, and Philipp
  Moritz.
\newblock Trust region policy optimization.
\newblock In {\em Proceedings of the 32nd International Conference on Machine
  Learning}, 2015.

\bibitem{socially-compatible}
Mehmet~Fatih Ozkan and Yao Ma.
\newblock Socially compatible control design of automated vehicle in mixed
  traffic.
\newblock {\em IEEE Control Systems Letters (L-CSS)}, 2021.

\end{thebibliography}

\end{document}